\documentclass[%
 reprint,
 amsmath,amssymb,
 aps,
 prapplied,
]{revtex4-2}

\usepackage{xcolor}

\usepackage{graphicx}
\usepackage{dcolumn}
\usepackage{bm}

\begin{document}

\preprint{APS/123-QED}

\title{The direction of the microjet produced by the collapse of a cavitation bubble located in between a wall and a free surface}

\author{Akihito Kiyama}
\affiliation{Institute of Global Innovation Research, Tokyo University of Agriculture and Technology, Japan}
\affiliation{Department of Mechanical and Aerospace Engineering, Utah State University, Logan, UT 84322, USA}
\author{Takaaki Shimazaki}
\affiliation{Department of Mechanical Systems Engineering, Tokyo University of Agriculture and Technology, Japan}
\author{Jos\'{e} Manuel Gordillo}
\affiliation{\'{A}rea de Mec\'{a}nica de Fluidos, Departamento de Ingenier\'{i}a Aeroespacial y Mec\'{a}nica de Fluidos, Universidad de Sevilla, Avenida de los Descubrimientos s/n 41092, Sevilla, Spain}
\author{Yoshiyuki Tagawa}
\affiliation{Institute of Global Innovation Research, Tokyo University of Agriculture and Technology, Japan}
\affiliation{Department of Mechanical Systems Engineering, Tokyo University of Agriculture and Technology, Japan}
\email{tagawayo@tuat.ac.jp}

\date{\today}

\begin{abstract}
In this paper, we present a simplified theoretical model based on the method of images that predicts the direction of the microjet produced after the implosion of the cavitation bubble created in between a free interface and rigid wall.
Our theoretical predictions have been verified by means of a thorough experimental study in which the distances of the pulsed-laser cavitation bubble to the wall and the free surface are varied in a systematic manner.
In addition, we extend the predictions to arbitrary values of the corner angle, $\pi/(2n)$ with $n$ a natural number.
The present analytical solution will be useful in the design of new strategies aimed at preventing the damage caused by cavitating bubbles over solid substrates.
\end{abstract}

\maketitle


\section{Introduction}

Cavitation bubbles can form a microjet directed towards a rigid wall when the collapse takes place near the solid surface \cite{Plesset1971}.
The impingement of the microjet on the solid boundary is one of the source of the cavitation-induced damage (e.g., \cite{Hsiao2014}).
A number of contributions have been devoted to predict and analyze the direction of the high-speed microjet
with different geometries or properties, such as the rigid flat surface \cite{Brujan2002}, the rigid curved surface \cite{Tomita2002}, the ice surface \cite{Cui2018}, the elastic surface \cite{Brujan2001}, the composite surface \cite{Tomita2003} and the free surface \cite{Gregorcic2007}, and they all reveal that the jet direction is  largely affected by the rigidity and the geometry of the flow boundaries.

Kucera and Blake \cite{Kucera1990} have studied the collapse of a cavitation bubble placed at the corner of two solid substrates forming an angle of either $\alpha=\pi/2$ or $\alpha=\pi/4$.
Brujan et al. \cite{Brujan2018} experimentally showed the direction of the microjet at the corner of two solid plates ($\alpha=\pi/2$) is proportional to the ratio of distances from the plates to the bubble. Tagawa and Peters \cite{Tagawa2018} employed the mirrored-image approach for solving the case $\alpha=\pi/n$, where $n$ is a natural number, and obtained analytical solutions for the jet direction that were confirmed experimentally. 
Recently, Wang et al. \cite{Wang2020} conducted numerical simulations that show a fair agreement with the published data for $\alpha=\pi/2$ \cite{Brujan2018} and their experiments for $\alpha=\pi/4$. The microjet direction inside the confined geometries \cite{Brujan2018b,Molefe2019} and that near the slot-like structured surface \cite{Andrews2020} have also been studied.

\begin{figure}
    \centering
    \includegraphics[width=0.9\columnwidth]{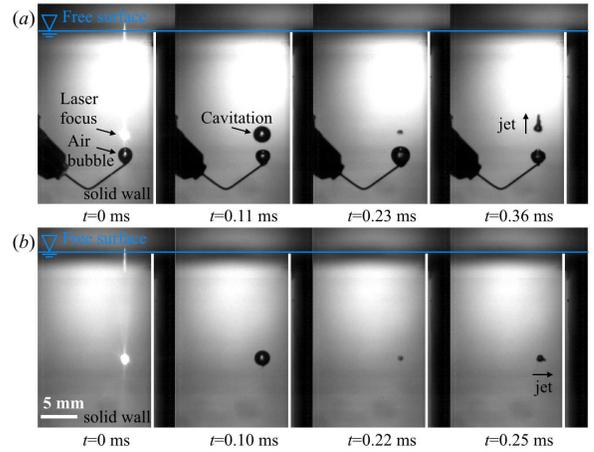}
    \caption{High-speed images of a collapsing bubble with a maximum diameter $D$ in the vicinity of a solid surface and of an air bubble ({\it a}) and that in the absence of the air bubble ({\it b}). The two cases shown have similar values of the ratios $w/D$ and $h/D$, with $w$ and $h$ denoting, respectively, the distances of the center of the cavitation bubble to the wall and the free surface.}
    \label{fig:snap2}
\end{figure}

In applications, it is desirable to prevent the impingement of the jet over the solid since, in this way, the lifetime of different types of fluid machinery such as ship propellers, can be substantially increased \cite{Brennen1995}.
An approach to achieve this is the installation of another surface with a different impedance.
Luo, et al \cite{Luo2019} placed an air bubble right next to the cavitation bubble and found that the direction of bubble movement upon its collapse was altered.
Indeed, our preliminary observation (Figure \ref{fig:snap2}({\it a})) shows that the jet produced by the implosion of a laser-induced cavitation bubble generated in the vicinity of an air bubble is directed away from it.
However, to our knowledge, the theoretical prediction of the microjet direction under the influence of both the free and rigid surfaces has not been provided yet. 

Here we provide a theoretical prediction of the microjet direction in between a wall and free surface as sketched in Figure \ref{fig:system}, which contains the major physical ingredients behind the control of the direction of the microjet by means of the addition of air bubbles near the collapsing vapor bubble.
Making use of ideas in a previous contribution \cite{Tagawa2018}, we employ the potential flow approximation and theoretically derive the direction of the jet as a function of the bubble location (Section II).
To make the derivation as simple as possible and without loss of generality, this model is particularized for the simplest case where the corner angle between a wall and free surface $\alpha=\pi/2(=90^\circ)$. 
Our analytical predictions are compared with the results of a systematic experimental study in which a laser-induced cavitation bubble is generated at well-defined distances from the wall and the free surface (Sections III \& IV).
In addition, we derive the generalized model for the jet angle $\beta$ for corner angles $\alpha=\pi/(2n)$ with $n$ a natural number (Section V).

\section{Theory}

\begin{figure}
    \centering
    \includegraphics[width=0.95\columnwidth]{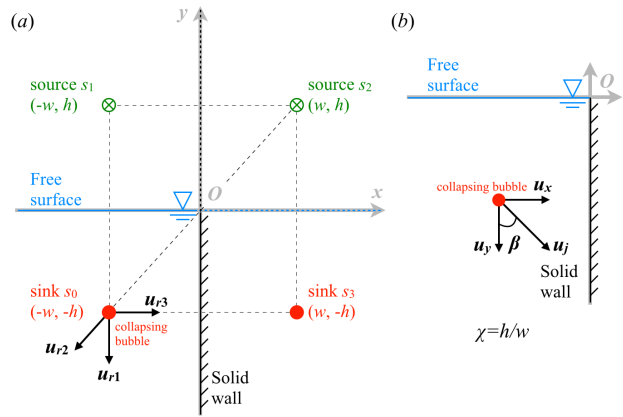}
    \caption{({\it a}) The schematic of the system on $x-y$ coordinates marked in grey. Green markers ($\bigotimes$) and red markers ($\bigcirc$) respectively represent the sources ($s_1 \& s_2$) and sinks ($s_0 \& s_3$). The velocity vectors affecting the bubble induced from mirrored images are denoted as $u_{r1}, u_{r2}$ and $u_{r3}$. ({\it b}) The illustration of the microjet direction $\beta$. The horizontal ($u_x$) and vertical ($u_y$) components of the velocity vector acting on the bubble $u_j$ are denoted.}
        \label{fig:system}
\end{figure}

A flow field induced by a spherically collapsing bubble is obtained by considering a sink \cite{Blake1987} (Figure \ref{fig:system}({\it a}), $s_0$ at ($-w, -h$)).
A solid wall and free surface are respectively modeled by placing the sink (Figure \ref{fig:system}({\it a}), $s_3$ at ($w, -h$)) and source (Figure \ref{fig:system}({\it a}), $s_1$ at ($-w, h$)) \cite{Blake1982}.
This sink-source system is appropriate since the bubble expansion does not significantly deform the boundaries in our experiments.
Noteworthy, another source $s_2$ must be placed at ($w, h$) to satisfy the boundary conditions as Tagawa \& Peters \cite{Tagawa2018} found the microjet direction $\beta$ (Figure \ref{fig:system}({\it b})) could not be modeled by simply considering the superposition of only two mirror images. 

We now derive the microjet direction $\beta$.
The radial speed $u_r$ at the certain distance $R$ from the sink/source is generally described as $u_r=Q/(4\pi R^2)$, where $Q$ is the flow rate.
The magnitude of the velocity vectors at the bubble location $u_{r1}, u_{r2}$ and $u_{r3}$ originated from respectively the sink/source $s_1, s_2$ and $s_3$ are
\begin{equation}
    u_{r1}=\frac{Q}{16\pi h^2}, u_{r2}=\frac{Q}{16\pi(h^2+w^2)},
    u_{r3}=\frac{Q}{16\pi w^2}.
\end{equation}
The microjet direction $\beta$ can be calculated as
\begin{equation}
    \beta=\arctan{\frac{u_x}{u_y}}=\arctan{\frac{1-(1+\chi^2)^{-3/2}}{\chi^{-2}+\chi(1+\chi^2)^{-3/2}}},
    \label{eq:model}
\end{equation}
where $u_x$ and $u_y$ are the magnitude of both the horizontal and the vertical components of the flow velocity vector as a function of $u_{r1}, u_{r2}, u_{r3}$ and the dimensionless bubble location parameter $\chi=h/w$ expressed as  
\footnotesize
\begin{eqnarray}
    u_x=u_{r3}-u_{r2}\frac{w}{\sqrt{h^2+w^2}}=\frac{Q}{16\pi w^2}\bigg[1-\frac{1}{(1+\chi^2)^{3/2}}\bigg], \\
    u_y=u_{r1}+u_{r2}\frac{h}{\sqrt{h^2+w^2}}=\frac{Q}{16\pi w^2}\bigg[\frac{1}{\chi^{2}}+\frac{\chi}{(1+\chi^2)^{3/2}}\bigg].
\end{eqnarray}
\normalsize
Equation \ref{eq:model} indicates that the direction of the microjet only depends on the dimensionless bubble location $\chi$.
For the extreme conditions, the model predicts $\lim_{\chi\rightarrow\infty}\beta=90^\circ$ and $\lim_{\chi\rightarrow0}\beta=0^\circ$.
We note that in contrast with the case of two rigid walls, where $\beta=45^\circ$ for $\chi=1$ \cite{Tagawa2018}, in the case at hand the deflection angle is $\beta=45^\circ$ for $\chi\simeq1.38$ by virtue of Equation 2.

\section{Experimental set-up}

\begin{figure}
    \centering
    \includegraphics[width=0.85\columnwidth]{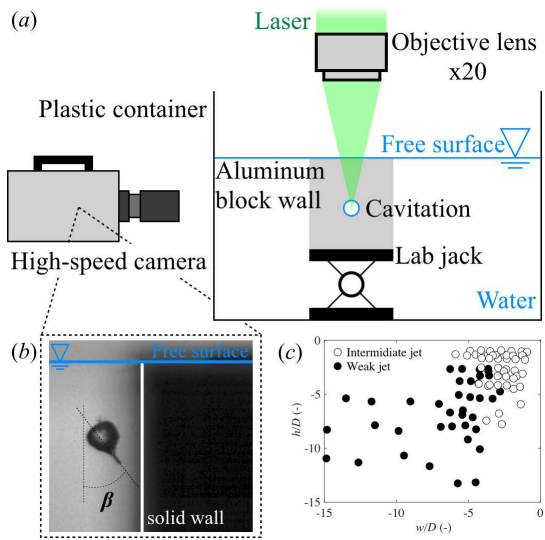}
    \caption{({\it a}) Experimental setup (not to scale). An aluminum block is used as a wall. The edge of the aluminum block reaches the free surface. A high-speed camera films the laser-induced cavitation bubble by using the backlight method. ({\it b}) A typical filmed image with the illustration of the angle $\beta$. ({\it c}) Parameter space, where $h/D=0$ and $w/D=0$ respectively indicate a free surface and wall.}
    \label{fig:setup}
\end{figure}

Figure \ref{fig:setup}({\it a}) shows the experimental set-up we used.
A pulsed laser beam (Nd:YAG laser Nano S PIV, Litron Lasers ltd., UK, wavelength 532 nm, pulse duration 6 ns) passes through the objective lens (MPLN series, Olympus Co., magnification 20 times, NA value 0.25) and illuminates a point inside ultra-purified water (produced by Milli-Q) inside an acrylic container (130 mm$\times$165 mm$\times$127 mm). 
A cavitation bubble emerges and expands.
A microjet then forms when the bubble collapses near the surface.
For the details of the laser and optical components, see \cite{Tagawa2016}.

A high-speed camera (FASTCAM SA-X, Photron Co., typical resolutions: 100,000 f.p.s. and $<$0.1 mm/pixel) is synchronized with the laser pulse through a delay generator (Model 575 Pulse/Delay Generator, BMC).
We controlled the maximum bubble diameter $D$, and the bubble location ($\chi=h/w$), and measured the direction of the microjet $\beta$ (see Figure \ref{fig:setup}({\it b})).
Experiments are repeated five times under each condition.

We observed the intermediate-jet (open markers in Figure \ref{fig:setup}({\it c}), $D=2.98\pm$0.46 mm) and the weak-jet (filled markers in Figure \ref{fig:setup}({\it c}), $D=1.39\pm$0.98 mm) based on the classification \cite{Supponen2016}.
Our experiments reveal that bubbles collapse preserving the spherical symmetry until the very end, when jet is issued.
We only analyzed the experimental data corresponding to the intermediate-jet regime because the jet is stronger, causes more severe damage, and it is much easier to identify experimentally.
The weak-jet was observed when the bubble is located far enough from surfaces, or the bubble diameter is small enough (Figure \ref{fig:setup}({\it c})), as expected. 

\section{Results}

\begin{figure}
    \centering
    \includegraphics[width=0.9\columnwidth]{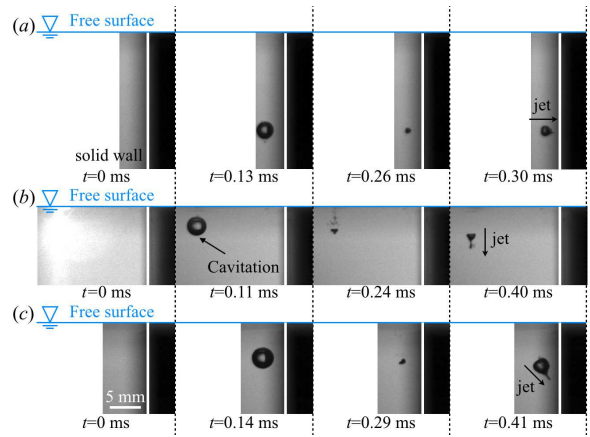}
    \caption{High-speed images of collapsing cavitation bubbles. ({\it a}) the microjet goes towards the solid wall ($\beta\simeq81^\circ$) where $\chi\simeq4.27$. ({\it b}) $\beta\simeq1^\circ$ for $\chi\simeq0.23$. ({\it c}) $\beta\simeq45^\circ$ for $\chi\simeq1.51$.}
    \label{fig:snap1}
\end{figure}

\begin{figure}
    \centering
    \includegraphics[width=0.9\columnwidth]{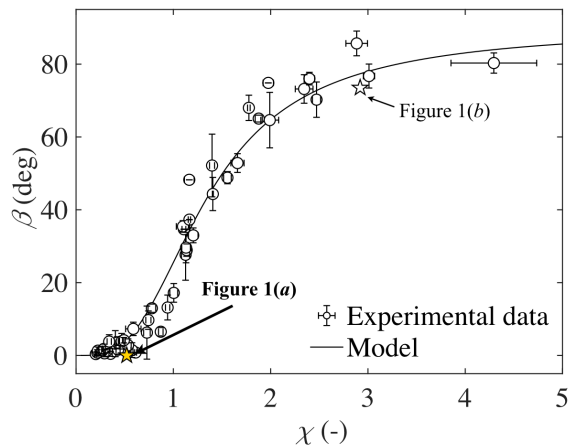}
    \caption{A comparison with the experimental data (circles) and the model (black line, Equation \ref{eq:model}) as a function of $\chi$. Markers and error bars respectively show the mean value of five runs and the standard deviation. The filled and open stars represent the values estimated from Figure \ref{fig:snap2}.}
        \label{fig:results}
\end{figure}

The typical behaviors of the cavitation bubble near the surfaces are shown in Figure \ref{fig:snap1}.
The microjet caused by the collapse of a bubble located near the wall ($\chi\simeq4.27$) is directed towards the solid substrate, with $\beta\simeq81^\circ$. In contrast, the bubble situated near the free surface ($\chi\simeq0.23$) produces a nearly vertical jet, with $\beta\simeq1^\circ$.
When the bubble is in between the rigid wall and free surface, with $\chi\simeq 1.5$, the deflection angle is $\beta\simeq 45^\circ$.

Figure \ref{fig:results} summarizes the experimental data and compares the measured jet angle $\beta$ versus $\chi=h/w$ with the values predicted by Equation \ref{eq:model}, finding a good agreement between observations and predictions.
The solid line denotes the prediction (Equation \ref{eq:model}), while the circles represent experimental results (Figure \ref{fig:setup}).
Overall, most of the data collapse into the solid line, which is insensitive to the values of the bubble diameters investigated herein.
Our model, which is a function of only the bubble location parameter $\chi=h/w$, accurately predicts the general trend of the microjet direction $\beta$ that occurs upon the collapse of a cavitation bubble affected by both the free surface and the wall.
Note that the model tends to overestimate $\beta$ slightly at the smaller $\chi$ values ($\chi<1$).
The model shows a steep rise of predicted $\beta$ values as a function of $\chi$ in that regime as the contribution of the sink $s_3$ will exceed those of sources $s_1$ and $s_2$.
The parameter $\chi$, which is a measure of the influence of two boundaries, might be too simple to describe $\beta$ perfectly in that regime.

Figure \ref{fig:results} also shows the results from Figure \ref{fig:snap2} using stars symbols.
The bubble shown in Figure \ref{fig:snap2}({\it b}), located at $\chi\simeq 2.9$, produces a jet with $\beta\simeq 73^\circ$, a value which agrees well with the prediction in Equation \ref{eq:model} (see an open star in Figure \ref{fig:results}).
To define $\chi$ for the data shown in Figure \ref{fig:snap2}({\it a}), we took the distance between the cavitation bubble center and the surface of the air bubble, $\lambda$, as the characteristic length, instead of $h$, where we find $\chi=\lambda/w\simeq$0.52.
Since the jet direction in Figure \ref{fig:snap2}({\it a}) seems to be almost vertical, we consider $\beta\simeq0^\circ$ (see a filled star in Figure \ref{fig:results}).
The fair agreement suggests that our model could even be applied to the situation where the jet direction is modified by an air bubble instead of by the free surface.

\section{An extension to various $\alpha$ values}

The approach of Tagawa \& Peters (2018) \cite{Tagawa2018} allows us to generalize the model for various $\alpha (=\pi/(2n))$.
As mentioned in the literature \cite{Tagawa2018}, the magnitude of the velocity vector $\vec{u_{s}}$ at the bubble located at $\vec{r_{s0}}=le^{i\theta_{s0}}$ induced by, for example a sink, located at $\vec{r_s}=le^{i\theta_{s}}$ is expressed as $|\vec{u_{s}}|=Q/(4\pi l^2)=Q/(8\pi l^2[1-\cos{(\theta_{s0}-\theta_{s})}])$, where $l$ is the distance between the sink and the center of the system $O$.
The unit vector for the sink with respect to the bubble is expressed as $(e^{i\theta_{s}}-e^{i\theta_{s0}})/(\sqrt{2[1-\cos{(\theta_{s0}-\theta_{s})}]})$.
The general expression of the velocity vector can be obtained by multiplying those relationships as

\begin{equation}
\vec{u_{s}}=\frac{Q}{\pi\sqrt{128}l^2}\frac{e^{i\theta_{s}}-e^{i\theta_{s0}}}{[1-\cos{(\theta_{s0}-\theta_{s})}]^{3/2}}.
\end{equation}

\begin{figure}
    \centering
    \includegraphics[width=0.95\columnwidth]{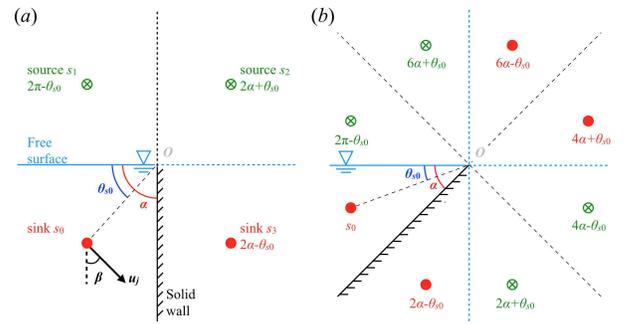}
    \caption{The schematic of the system on the polar coordinates. Green markers ($\bigotimes$) and red markers ($\bigcirc$) respectively represent the sources and sinks, where $s_0$ represents the cavitation bubble. The corner opening angle $\alpha$ is $\pi/2(=90^\circ)$ in ({\it a}) and $\pi/4(=45^\circ)$ in ({\it b}). The sinks/sources in ({\it a}) are named as $s_1, s_2, \& s_3$ for comparison purposes to figure \ref{fig:system}. The angle of image sinks/sources is denoted as a function of both the corner opening angle $\alpha$ and the bubble location $\theta_{s0}$.}
        \label{fig:system2}
\end{figure}

The system for $n=1$ considered here (Figure \ref{fig:system2}({\it a})) corresponds to that shown in Figure \ref{fig:system}.
By applying the same strategy, we may place the sinks/sources for $n=2$ (i.e., $\alpha=\pi/4(=45^\circ)$) as shown in Figure \ref{fig:system2}({\it b}), or even higher $n$ values ($\alpha=\pi/(2n)$). 
The vector for the microjet $\vec{u_j}$ is given by summing up all $\vec{u_s}$ belong to the image sinks/sources, as

\begin{equation}
\footnotesize
\begin{split}
\frac{\vec{u_{j}}}{C}&=\frac{e^{i\theta_{s0}}-e^{i(2\pi-\theta_{s0})}}{[1-\cos{(2\theta_{s0})}]^{3/2}}\\&\quad+\sum_{k=1}^{2n-1}\bigg[\frac{e^{i\theta_{s0}}-e^{i(2k\alpha-\theta_{s0})}}{[1-\cos{(2\theta_{s0}-2k\alpha)}]^{3/2}}-\frac{e^{i\theta_{s_0}}-e^{i(2k\alpha+\theta_{s0})}}{[1-\cos{(2k\alpha)}]^{3/2}}\bigg](-1)^k.
\end{split}
\normalsize
\end{equation}

The first term represents the influence of the induced flow by a source located at $\theta_{s}=2\pi-\theta_{s0}$, while the latter terms indicate those by the sinks/sources ($n-1$ for each) located at $\theta_{s}=2k\alpha\pm\theta_{s0}$.
Note that we assumed that all sinks/sources are located at the same distance, $l$, from the center of the system $O$.
The natural number $k$ represents the index from 1 to $2n-1$.
The prefactor, $C$, in the equation above is expressed as $C=Q/(\pi\sqrt{128}l^2)$.
The jet direction is thus obtained as
\begin{equation}
\beta=\arctan{\bigg(-\frac{\mathrm{Re}(\vec{u_j})}{\mathrm{Im}(\vec{u_j})}\bigg)},
\label{eq:model2}
\end{equation}
without using prefactor $C$.

Figure \ref{fig:system2} shows the normalized jet angle, $\beta/\alpha$, as a function of the location of the bubble, $\theta_{s0}/\alpha$, predicted by Equation \ref{eq:model2}.
We note that, for $n=1$ (i.e., $\alpha=\pi/2 (=90^\circ)$), the prediction matches with Equation \ref{eq:model} (not shown).
The well-defined non-dimensional axes reveal the fact that increasing $n$ values (i.e., the finer corner angle $\alpha$) induces a slight shift of the curve as seen in the inset. 
However, even though we increase $n$ significantly, the influence of the first term in Equation \ref{eq:model2} remains in effect and the curve does not cross the point $\beta/\alpha=\theta_{s0}/\alpha=0.5$.
It perhaps belongs to the same reason as to why Equation \ref{eq:model} predicts $\beta=45^\circ$ at not $\chi=1$ but $\chi\simeq1.38$, meaning that this is a unique feature of such a system made by two different boundaries; the free surface has a relatively dominant effect on the jet direction, $\beta$, when compared to the solid wall.
Though the situation for $\alpha<\pi/2$ is a bit difficult to test experimentally and thus speculative, our extension of the previous work \cite{Tagawa2018} to our system might deepen our understanding of the jet formation upon the cavity collapse at the complex geometry.

\begin{figure}
    \centering
    \includegraphics[width=0.95\columnwidth]{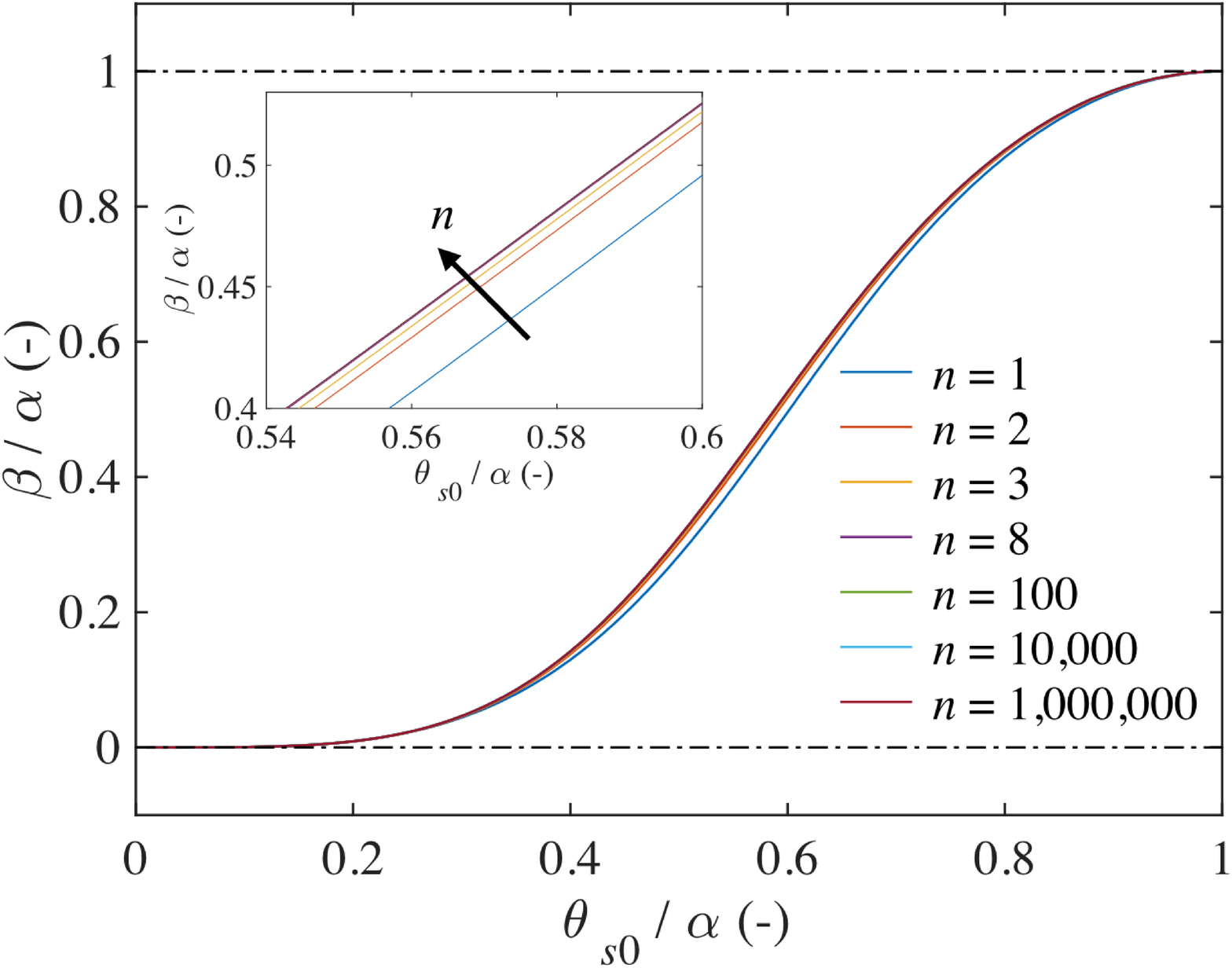}
    \caption{The predictions of the jet angle $\beta/\alpha$ as a function of the bubble position $\theta_b/\alpha$ calculated by equation \ref{eq:model2} with varying $n$ values. A zoom-in view is shown in the inset. Dashed lines indicate $\beta/\alpha=0$ and $\beta/\alpha=1.0$.}
        \label{fig:result2}
\end{figure}

\section{Conclusion}

We investigated the direction $\beta$ of the microjet caused by a cavitation bubble collapsing in between a wall and free surface.
We derived the theoretical prediction for the jet angle deflection, $\beta$, as a function of the bubble location parameter $\chi=h/w$ (Equation \ref{eq:model}) based on the sink-source system (Figure \ref{fig:system}).
We compared the model predictions with our systematic experiment (Figure \ref{fig:snap1}), finding a fair agreement between the calculated and measured values (see Figure \ref{fig:results}).
The model is even potentially applicable to the case where the jet is influenced by an air bubble (Figure \ref{fig:snap2}).
We note that the model might not be applicable to describe the cases when the air bubble is located in between the cavitation bubble and the wall, or the wall is covered by bubbles \cite{Avila2020}.
In addition, we extended the approach presented by Tagawa \& Peters (2018) \cite{Tagawa2018} to our system and derived the equation that predicts the jet direction, $\beta$, caused by the bubble at a corner with of angle, $\alpha=\pi/(2n)$ (see Equation \ref{eq:model2}), where the case for $n=1$ leads the same result as that from Equation \ref{eq:model}.
A numerical calculation with sufficiently large $n$ values revealed that the free surface has a relatively dominant effect on the microjet direction upon the cavitation collapse when compared to the solid wall (Figure \ref{fig:result2}).

\section{Acknowledgments}

We thank Mr. Takuya Yamaguchi for his support in the experiments.
A. K. pursued the study first while affiliated at Tokyo University of Agriculture and Technology.
A.K. is now at Utah State University as a JSPS Overseas Research Fellow.
This project is supported by the Institute of Global Innovation Research in TUAT for J.M.G.’s summer stay at TUAT.
Y.T. acknowledges financial support from JSPS KAKENHI Grant Nos. 17H01246 and 20H00223.

\bibliography{Aki}

\end{document}